\documentclass[amsmath,amssymb,prb,twocolumn,notitlepage,showpacs,unsortedaddress,aps]{revtex4-1}
\usepackage{graphicx}

\newcommand\be{\begin{equation}}
\newcommand\ee{\end{equation}}

\newcommand{\tot}{{\rm tot}}

\begin{document}

\title{Spectral gaps of AKLT Hamiltonians using Tensor Network methods}
\author{Artur Garcia-Saez}
\affiliation{C.N. Yang Institute for Theoretical Physics, State University of New York, Stony Brook, New York, USA}
\author{Valentin Murg}
\affiliation{Institute for Theoretical Physics, University of Vienna, Vienna, Austria}
\author{Tzu-Chieh Wei}
\affiliation{C.N. Yang Institute for Theoretical Physics, State University of New York, Stony Brook, New York, USA}

\begin{abstract}
Using exact diagonalization and tensor network techniques we compute
the gap for the AKLT Hamiltonian in 1D and 2D spatial dimensions.
Tensor Network methods are used to extract physical properties
directly in the thermodynamic limit, and we support these results
using finite-size scalings from exact diagonalization. Studying the
AKLT Hamiltonian perturbed by an external field, we show how to
obtain an accurate value of the gap of the original AKLT Hamiltonian
from the field value at which the ground state verifies $e_0<0$,
which is a quantum critical point. With the Tensor Network
Renormalization Group methods we provide direct evidence of a finite gap
 in the thermodynamic
limit for the AKLT models in the 1D chain and 2D hexagonal and
square lattices. This method can be applied generally to
Hamiltonians with rotational symmetry, and we also show results
beyond the AKLT model.
\end{abstract}

\pacs{0.00}

\maketitle

\section{Introduction} \label{section:introduction}

In quantum many-body systems, the low energy behavior is determined
by the ground state and low lying excitations. When there is a
spectral gap above the ground state, the system is usually robust
against small perturbations. Among the well-known examples are the
superconductivity and integer and fractional quantum Hall effects.
The existence of a spectral gap above the ground state is also an
important condition for robust topological phases, either intrinsic
or symmetry-protected. In one dimensional spin chains the relation
of the Hamiltonian and the existence of a spectral gap has been
extensively studied and understood. For example,  Haldane provided
convincing field-theory arguments on the existence of a finite
spectral gap for integer-spin Heisenberg
chains~\cite{Haldane,Haldane2}, in contrast to half-odd-integer
spins~\cite{LSM}. This conjecture was later substantiated by a
rotationally invariant model of a spin-1 chain constructed by
Affleck, Kennedy, Lieb and Tasaki (AKLT), in which an exact ground
state wave function is known and the finite spectral gap can be
shown to exist. Their construction was generalized and
extended~\cite{AKLT2,FannesNachtergaeleWerner92}, and in particular
the technique of establishing the spectral
gap~\cite{Knabe,Nachtergaele}.


In addition to the one-dimensional spin-1 model, AKLT also
generalized their valence-bond solid (VBS) construction to higher
dimensions~\cite{AKLT2}. The AKLT Hamiltonians involve only
nearest-neighbor two-body interactions and posssess the rotational
symmetry of spins and the spatial symmetry of the underlying
lattice. With suitable boundary conditions, their ground states are
unique and respect the symmetries of the Hamiltonians. However, the
existence of the spectral gap above the unique ground state has not
been established rigorously. What was known is that the correlations
are decaying exponentially, e.g., for the AKLT ground states on the
honeycomb and square lattice~\cite{AKLT2}. This only suggests that
the gap is likely to exist, since it is neither known nor
necessary that exponential-decay correlation functions imply the
existence of a gap, unless the system has Lorentz symmetry. However, 
if the system has a finite gap above its ground state
the connected correlation functions are known to decay
exponentially. It is also known~\cite{PowerLaw} that if the
ground-state correlation functions have power-law decay, the system
is gapless~\cite{PowerLaw,PowerLaw2}.
These intuitions are reinforced by numerical studies of the AKLT in the 
honeycomb lattice, first performed by Ganesh et al., and the finite size 
scaling\cite{ganesh} shows a robust finite spin gap.

In a very different research direction, AKLT states have recently been
explored in the context of quantum computation by local
measurement~\cite{Oneway,Oneway2,RaussendorfWei12}.
In particular, the spin-1 AKLT
chain can be used to simulate single-qubit gate operations on a
single qubit~\cite{Gross,Brennen}, and the spin-3/2 two-dimensional
AKLT state on the honeycomb lattice can be used as a universal
resource~\cite{WeiAffleckRaussendorf11,Miyake11,WeiAffleckRaussendorf12}.
 An important question regarding universal  resource states is whether they
can be the unique ground state of a physically reasonable, gapped
Hamiltonian~\cite{Nielsen06}. If so, these quantum computational
resource states may be created by quantum engineering the
Hamiltonian and cooling the system to sufficiently low temperature.
Therefore, the issue of spectral gaps in AKLT Hamiltonians in two
and higher dimensions becomes even more pressing.

\begin{center}
\begin{figure}[htb]
\includegraphics[width=0.9\linewidth]{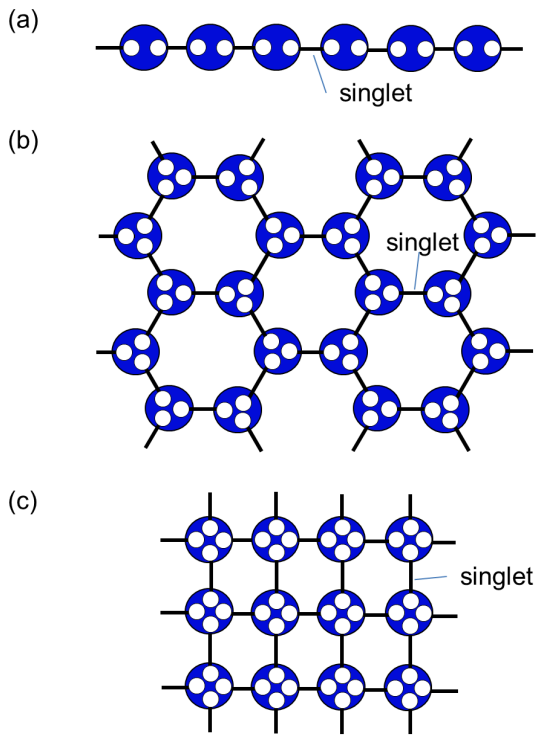}
\caption{The Valence Bond Solid (VBS) is constructed by allocating
on each site as many spin-1/2 particles (in their symmetric
subspace) as neighboring sites, e.g., (a)  2 for 1D chains, (b) 3
for the hexagonal lattice, and (c) 4 in the 2D square lattice. For
each pair of neighboring sites, the VBS state places a singlet
between two spin-1/2. Having a singlet connecting sites imposes
restrictions on the total spin evaluated on neighboring sites by
$P^{(S=s_{\max})}$.}\label{fig:aklt}
\end{figure}
\end{center}

Here, we study the spectral properties of the AKLT Hamiltonians, and
in particular, we investigate the energy gap of 1D spin-1 chains and
2D hexagonal and square lattices. The general formulation of the
AKLT Hamiltonian is a set of projectors acting on nearest neighbors,
and projecting into the subspace with maximum spin magnitude that
two neighboring can in principle form:\be H_{\rm AKLT} =
\sum_{\langle i,j\rangle} P^{(S_{ij}=s_{\max})}. \ee For example,
$s_{\max}=2$ for the spin-1 chain, $s_{\max}=3$ for the spin-3/2
AKLT model on the honeycomb lattice and $s_{\max}=4$ for the spin-2
model on the square lattice. Defined in this way, the Hamiltonian is
semidefinite positive, and the ground state has energy $E_0=0$. The
ground state of this system is a VBS~\cite{AKLT} (see
Fig.\ref{fig:aklt}), which is constructed by allocating on each site
as many (virtual) spin-1/2 particles as neighboring sites $z$ (e.g.,
$z=2$ for 1D chains, $z=3$ for the hexagonal lattice, $z=4$ in the
2D square lattice) in their symmetric subspace. The local spin
magnitude at each site is thus $z/2$. For each pair of neighboring
two sites, the VBS construction places a singlet between the
associated two virtual spin-1/2's. Having a singlet connecting sites
imposes restrictions on the total spin evaluated on neighboring
sites. In 1D, for example, two neighboring sites cannot add up to
spin 2, so the projector $P^{(S=2)}$ (more generally,
$P^{(S=s_{\max})}$) to this subspace will yield a zero value when
evaluated over the VBS. Being a semipositive definite operator, the
VBS is the ground state. With an appropriate boundary condition, it
is the unique ground state~\cite{KLT}.

Let us briefly explain why the AKLT Hamiltonians are spin rotation
invariant. Consider the spin-1 chain. Let $\vec{S}_{ij}\equiv
\vec{S}_i + \vec{S}_j$. Then projection of sites $i$ and $j$ to
their joint spin-2 subspace is given by $P=\lambda [S_{ij}^2 -
0][S_{ij}^2- 1(1+1)]$, as it annihilates states in $S_{ij}=0,1$
subspaces.  The coefficient $\lambda$ is determined by the
requirement that $P$ is a projection on the $S_{ij}=2$ subspace,
i.e., $P|S_{ij}=2,S_z\rangle=|S_{ij}=2,S_z\rangle$, which leads to
$1/\lambda=[2(2+1)-0][2(2+1)-1(1+1)]=12$. Then by expanding
$S_{ij}^2= S_i^2 +S_j^2 + 2\vec{S}_i\cdot \vec{S}_j$ and using that
$S_i^2$ and $S_j^2$ are constants (2 for spin 1) we obtain that the
Hamiltonian contains a polynomial of $\vec{S}_i\cdot \vec{S}_j$ up
to degree 2 (and $s_{\max}$ in general). The resultant AKLT
Hamitonians for the spin-1 chain, the spin-3/2 honeycomb lattice,
and the spin-2 square lattice are shown in Eqs.~(\ref{eqn:1DAKLT}),
(\ref{eqn:spin3/2AKLT}), and (\ref{eqn:spin2AKLT}), respectively.
Thus we arrive at the conclusion that the AKLT Hamiltonian is spin
rotation invariant and, additionally, semidefinite positive. Its
ground state, the AKLT wavefunction, has energy $ E_0= 0$. This
ground state --the VBS-- has also total angular momentum $S^{\rm
tot}=0$ and $z$ component $S^{\rm tot}_z=0$. The AKLT Hamiltonian
commutes with $(S^{\rm tot})^2$ (where $\vec{S}^{\rm tot}\equiv
\sum_i \vec{S}^{(i)}$) and $S^{\rm tot}_z\equiv\sum_i S^{(i)}_z$,
thus they are good quantum numbers and the eigenstates of the AKLT
(or any such spin-rotation invariant) Hamiltonian can be labeled by
$|S^{\rm tot},S^{\rm tot}_z,\alpha\rangle$, where $\alpha$ is some
additional labeling for different ways of constructing  $|S^{\rm
tot}, S^{\rm tot}_z\rangle$ states. Furthermore, for a given $S^{\rm
tot}$ and $\alpha$, the different sublevels characterized by $S^{\rm
tot}_z$ are degenerate.

We show here how the conservation of the angular momenta can be
utilized to compute the gap (even in the thermodynamic limit), by
adding a local field term $h\,S^{\rm tot}_z$ to the Hamiltonian. We
shall consider the total Hamiltonian \be \label{eq:Hfield} H(h) =
H_{\rm AKLT} + h \sum_i S^{(i)}_z. \ee Because of $[S^{\rm
tot}_z,H_{\rm AKLT}]=0$, the VBS ground state does not interact with
the local field and has ground state energy remaining zero for any
value of $h$. For $h>0$, any excited state with $S_z\neq 0$
interacts with the local field and the corresponding zero-field
degenerate levels split linearly with $S_z$. Here we assume that
spin triplets are elementary excitations, as will be justified
later. If there is a gap, for some field value $h_t>0$ the energy of
some level will cross the zero energy to negative, becoming the
ground state of the system at the field $h_t$. We can observe this
transition by computing the energy or the z-component total angular
momentum $S_z^{\rm tot}$ of the ground state and detecting the
transition to $S_z \neq 0$ (nonzero magnetization). Interestingly,
this can be computed efficiently in the TN description.

\begin{center}
\begin{figure}[htb]
\includegraphics[width=0.65\linewidth]{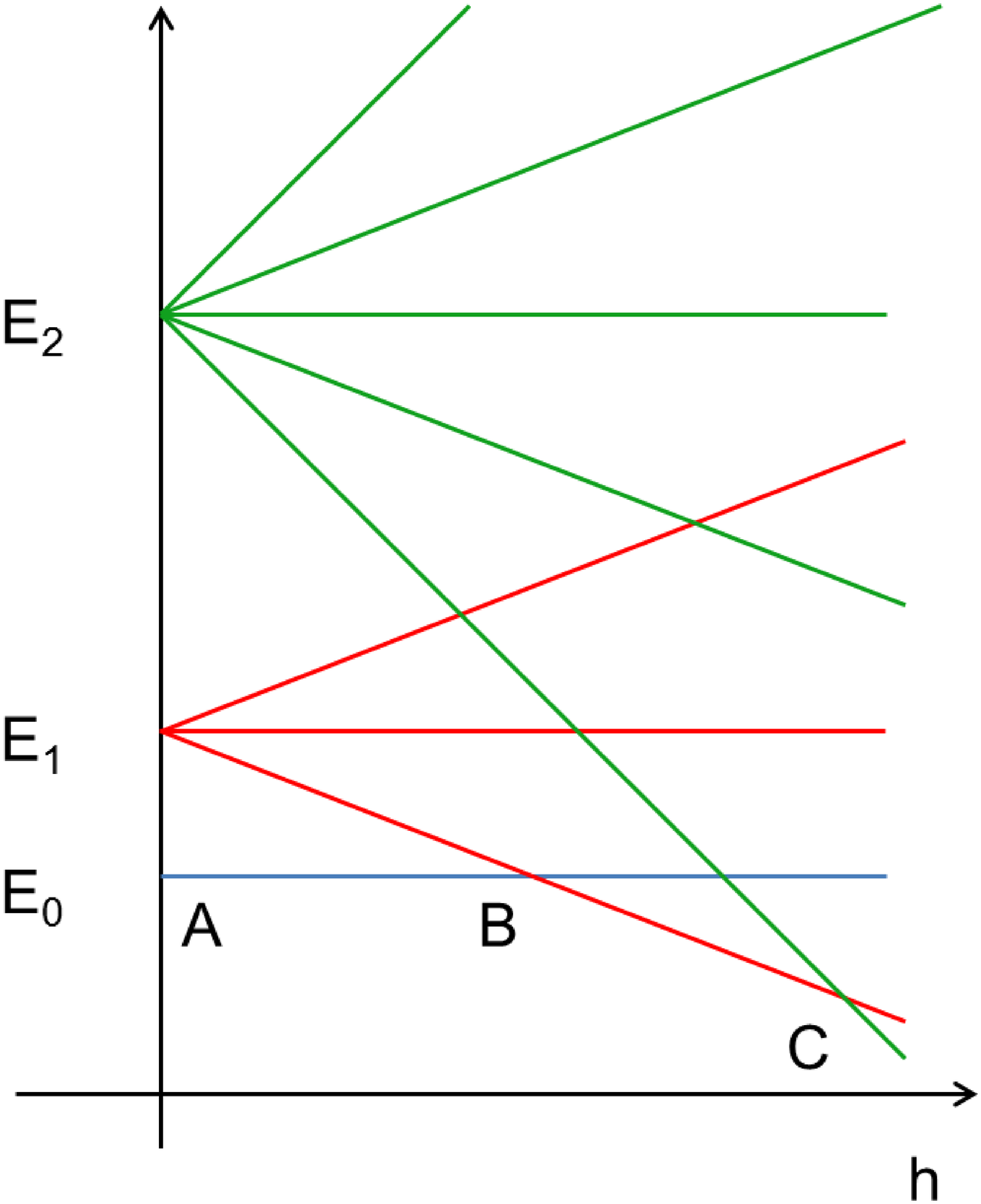}
\includegraphics[width=0.3\linewidth]{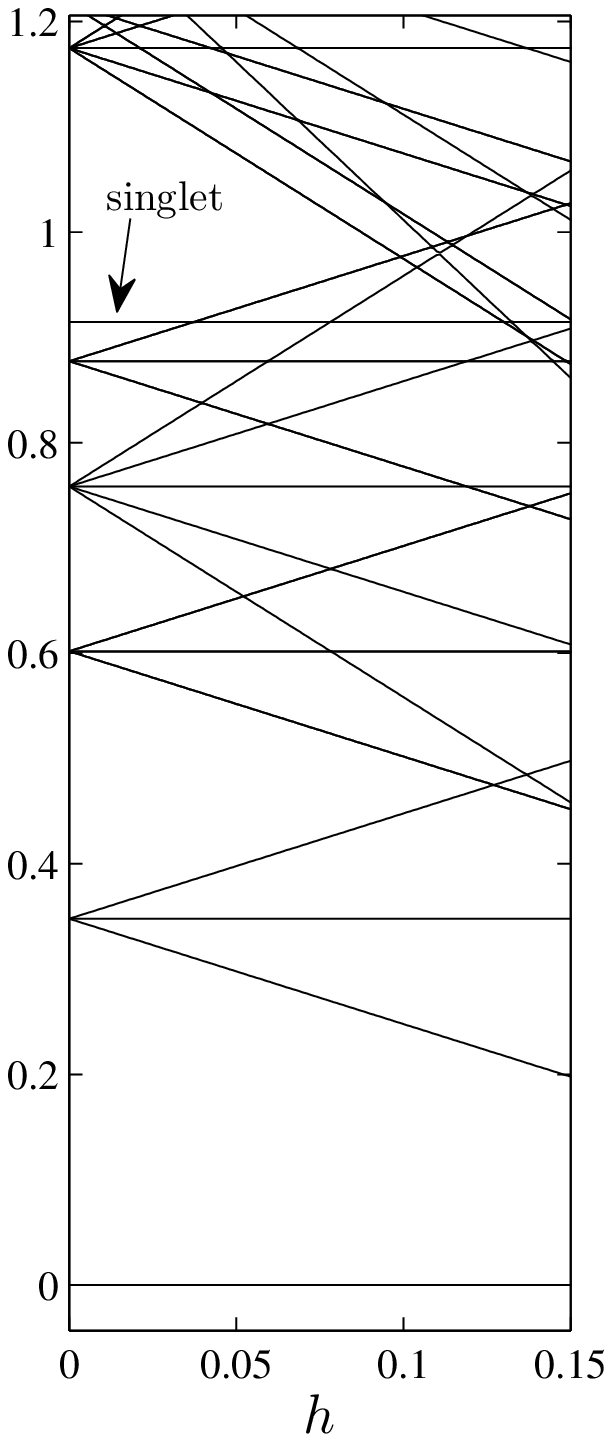}
\caption{Left: A schematic picture for the level splitting under the
external field $h$ for a generic antiferromagnetic spin-rotation
invariant Hamiltonian. A represents the location of the ground-state
energy at the zero field. The crossings at B, C, etc., represent the
ground state switching to one with a different set of quantum
numbers. In the thermodynamic limit, the envelope curve for the
$E_0$ plateau (if it exists) and the successive crossings form the
ground state energy curve for the system in presence of the external
field $h$. Knowing the first crossing and the slope, one can
extrapolate to obtain the zero-field spectral gap. If the first
crossing is by the one with a slope $-1$, then the value of $h$ at
the crossing point gives the spectral gap. One is interested in the
gap in the thermodynamic limit, i.e., the property of this crossing
in such limit. Right: The actual spectrum of a $N=8$
AKLT chain with external field (Eq.\ref{eq:Hfield}). Notice the presence
of singlets as excitations at high energies.}\label{fig:schematic}
\end{figure}
\end{center}

As shown schematically in Fig.~\ref{fig:schematic}, the response of
the non-zero angular momentum states is linear with the field, so
from the slope in the energy curve for $h>h_t$ we can determine (by
linear extrapolation) the energy of each state at $h=0$. As for the
gap, we are interested in the first excited state(s). It might be
possible that as the field is increased such states never appear as
the ground state (of the field-dependent Hamiltonian), as they may
be (i) insensitive to the field, i.e., they are also states
characterized by $S^{\rm tot}=S^{\rm tot}_z=0$, or (ii) before they
cross the zero energy curve, higher energy states already cross the
zero. We shall argue and provide evidence from exact diagonalization
on small systems that (i) the first excited states of
antiferromagnetic Hamiltonians (such as Heisenberg and AKLT) are
never characterized by $S^{\rm tot}=S^{\rm tot}_z=0$ but by $S^{\rm
tot}=1, S^{\rm tot}_z=\pm1,0$. If case (ii) occurs, then the
determination of the gap from our method will turn into a set of
lower bounds. However, we shall also show numerical evidence that
(ii) does not occur in our consideration. Moreover, analysis from
field-theory treatment on the spin-1 Heisenberg chain predicts that
triplets are indeed the lowest excitations \cite{affleckPRB}. An
important part of this work is dedicated to show that the identified
transition shows the exact value of the gap.

In this work we utilize various numerical methods, from exact
diagonalization and 1D Matrix Product States (MPS) variation method and iTEBD,
to the 2D Tensor Network Renormalization Group (TNRG).
These methods have nowadays become standard tools to obtain phyical properties
in the thermodynamic limit. In the following,
we only mention essence of these methods and we refer the readers to
the literature for detailed implementation.

We compute the gap of the AKLT Hamiltonian in 1D for finite chains of increasing length,
and we directly study as well the infinite case by means of MPS techniques.
The resulting gap value $\Delta\cong0.350$ agrees with previous
bounds, but under the above-mentioned assumptions our value is a
direct estimation of the gap. We apply the same ideas to the
spin-3/2 hexagonal lattice, and find a gap $\Delta\cong0.101$ using infinite-size Tensor Network methods. 
Our value is in excellent agreement 
with previous results obtained by exact diagonalization over small clusters\cite{ganesh} of $N=12-18$ spins.
We also obtain a value of the gap $\Delta\cong0.03$ for the spin-2
square lattice. Notice that all these gap values are obtained from
the AKLT Hamiltonian formulated to be the sum of projectors (see
Eq.\ref{eqn:1DAKLT}, \ref{eqn:spin3/2AKLT} and \ref{eqn:spin2AKLT}),
with a coupling term $J<1$ that rescales the energy spectrum,
compared to the corresponding Heisenberg Hamtiltonians.

We organize this paper showing first exact results for the 1D spin-1 AKLT model in Section
\ref{section:1Dexact}. We apply MPS techniques to compute the gap in the thermodynamic limit
in Section \ref{section:mps}, where we check our method gainst DMRG results for the spin-1
Heisenberg chain. Then we show how similar techniques can be used
for the computation of the gap in 2D systems: in Section \ref{section:hexagonal} we
show results for the hexagonal spin-3/2 lattice, and in Section
\ref{section:square} we compute the gap of the square spin-2 lattice.
Finally, conclusions are presented in Setion \ref{section:conclusions}.

\section{1D AKLT Hamiltonian} \label{section:1Dexact}

In this section we put onto solid grounds and elaborate the ideas
presented in the introduction by running some exact calculations for
small 1D systems. We explore the energy spectrum around $h=0$ for
the total Hamiltonian in Eq.~(\ref{eq:Hfield}) where for spin-1
systems we have \be \label{eqn:1DAKLT} H^{S=1}_{\rm AKLT}
=\frac{1}{2} \sum_{\langle i,j\rangle} \left[ {\vec{S}_i
\cdot\vec{S}_j} + \frac{1}{3}(\vec{S}_i\cdot\vec{S}_j)^2+
\frac{2}{3}\right] \ee We note that due to the requirement of being
a projector for nearest-neighbor interaction, there is a factor
$J=1/2$ compared to the usual Heisenberg Hamiltonian with $J=1$. In
order to ensure the uniqueness of the ground state in finite
calculations we impose periodic boundary conditions.

\begin{center}
\begin{figure}[htb]
\includegraphics[width=.9\linewidth]{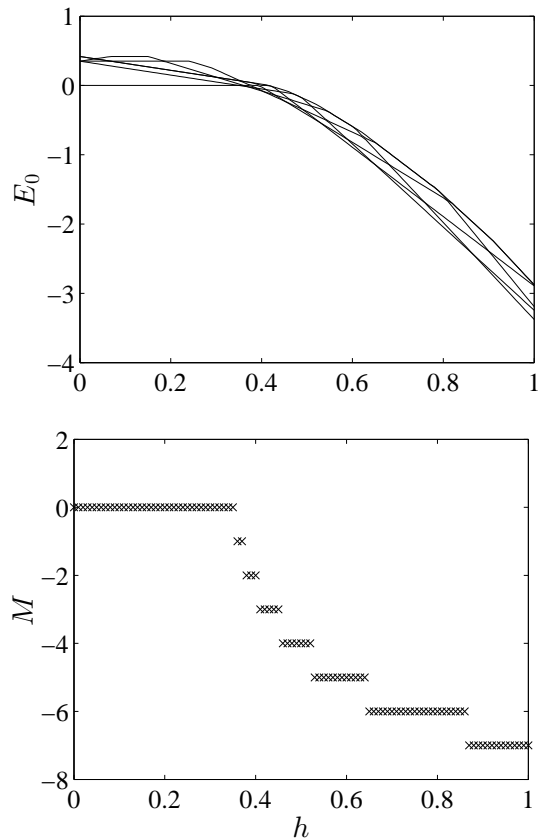}
\caption{For the Hamiltonian Eq.\ref{eq:Hfield} we plot
the energy per particle (top) of the $6$ eigenstates with lower energy, and $S_z^{tot}$
of the ground state (bottom) as a
function of the field $h$, with system size $N=12$ and PBC. At $h=0$ we observe a gap $\Delta$ between
the ground state and the triple-degenerate first excited state. Of these 3 sates,
one crosses the ground state at $h=\Delta$ after interacting linearly with the field.
At this point, the ground states makes a transition to $S_z^{tot} = -1$.
Successive crossings of higher $S_z^{tot}$ will further decrease the total $S_z^{tot}$
of the ground state an they become the new low energy states,
up to a total $S_z^{tot}=Ns$.} \label{fig:exact_12_detail}
\end{figure}
\end{center}

We compute the energy of the $6$ lowest energy states for different
values of $h$ and evaluate $S_z^{tot}$ for the ground state. In
Fig.\ref{fig:exact_12_detail} we plot these values for a system of
$N=12$ as an illustration. In the upper panel we plot the energy per
particle of each energy level, and the lower panel shows $S_z^{tot}$
of the ground state. At $h=0$ we observe a ground state with $E_0 =
0$ --the AKLT state-- and a gap with the first excited state with a
triple degeneracy and $S_z^{tot} = -1,0,1$. The fact that the energy
splits into three levels linearly with the field shows that they
have $S_{\rm tot}=1$. Of these 3 lowest-energy excited states, one
with $S_z^\tot = -1$ interacts with the field $h>0$ linearly with
slope $S_z^\tot = -1$. At a given value of $h_t =\Delta$ this
excited state has energy $E_1=0$, and becomes the ground state for
$h>h_t$. Other higher levels will come down and cross this level for
larger $h$. The original ground state, having $S_z^\tot= 0$ keeps
having $E_0=0$ for any finite $h$. In this scenario there are two
ways to extract the gap. First, we can extrapolate from the value of
$h=h_t$ back linearly to $h=0$ and locate the cross point with the
y-axis. Second, without extrapolation, this value of the field $h_t$
provides a direct reading of the exact value of the energy gap.
(This  second view  will be useful in the thermodynamic limit, where
one has only access to energy and magnetic moment per site). The
transition to a ground state with $S_z^{tot} = -1$ is just the first
of a series of successive transitions to ground states with
increasing $S_z^{tot}$ which appear after the crossing of the
current ground state with excited states with higher $S^{tot}$ --and
thus a steepest slope in the energy spectrum. Moreover, for each
value of $h$, $S_z^{tot}$ is the derivative of the energy and we can
observe the transition from these two magnitudes.

\begin{center}
\begin{table}
\begin{tabular}{ r | l | l || l}
 & \multicolumn{2}{c}{AKLT} & Heisenberg\\
N & $E_1-E_0$ & $E_2-E_0$ & $E_1 - E_0$\\
\hline
8 &  0.349849122 & 0.4988577 & 0.5935552\\
10 & 0.350091873 & 0.4477184 & 0.5248079\\
12 & 0.350120437 & 0.4187836 & 0.4841964\\
14 & 0.350123733 & 0.4009563 & 0.4589653\\
16 & 0.350124109 & 0.3892351 & 0.4427955\\
\end{tabular}
\caption{Exact gap calculation for spin chains in periodic boundary
conditions with increasing length $N$.
We show also the energy gap to the second excitation.}\label{fig:table_gap}
\end{table}
\end{center}

\begin{center}
\begin{figure}[htb]
\includegraphics[width=\linewidth]{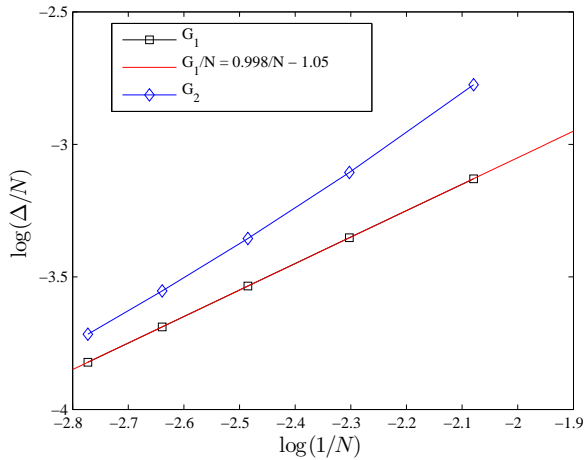}
\caption{From the exact diagonalization we extract estimations of
the gap in the thermodynamic limit of the AKLT Hamiltonian. Left:
The energy gap per site $\Delta/N$ vs. $\frac{1}{N}$ for $N=8,10,12,14,16$.
The slope of this plot, wtih a good linear fitting, provides the gap
estimation $\Delta = 0.350$. We plot in  blue the difference between the gap to the
first excitation and to the second excitation $(E_2-E_1)/N$ vs. $\frac{1}{N}$
for $N=8,10,12,14,16$. With a good square fitting, we observe at
$\frac{1}{N}\rightarrow 0$ we a small but positive value of $(E_2-E_1)/N$,
so we expect that the first transition to $S_z=-1$ is robust
in the thermodynamic limit.} \label{fig:finite_gap_scal}
\end{figure}
\end{center}

Our exact results for different values of $N$ are shown
in Table~\ref{fig:table_gap}, including similar calcuations for the Heisenberg model.
Unlike this latter case, in the AKLT we observe a growing energy gap for
longer systems, converging to a value $\Delta = 0.350$. This is in
agreement with previous upper bounds\cite{Knabe,Auerbach}. The
second transition converges to $\Delta_2 = \Delta$, which
suggests that even in the thermodynamic limit the first transition
to a $S_z^{tot}\neq0$ ground state will take place to a state with
$S_z^{tot}=1$. As shown in Fig.\ref{fig:finite_gap_scal} the gap for the AKLT
has a clear $\frac{1}{N}$ dependence. We will complete this result in the following sections with
direct numerical calculations  in the thermodynamic limit using
Tensor Network methods, but this finite size scaling
suggests that our readings of the transition in the ground state at
$h_t$ provide a direct measurement of the gap.

\section{MPS calculations of finite and infinite 1D systems} \label{section:mps}

The exact results in Section \ref{section:1Dexact} confirm the
picture described in Section \ref{section:introduction}, and show
how one can extract the  value of the gap from the magnetization of
ground states. Unfortunately by exact diagonalization, this can be
computed for only a few particles. We complete these calculations by
showing results computed using tensor network techniques in 1D
systems. The following sections will present results for 2D
lattices.

MPS and tensor networks provide a detailed description of ground
states of local Hamiltonians, and are efficient methods for systems
with an energy gap. This implies that for the gapped phase we are
exploring here, we can obtain good approximations to the ground
state at $h=0$ if there is a gap in the AKLT model. However, we have
seen that this gap closes for higher $h$, and then follows a
succession of transitions to new ground states. The region near the
first transition represents a region close to a continuous quantum
phase transition. This translates into a not-so-efficient
description of the ground state in this region, so we expect the
need of higher bond dimension $\chi$ for the ground state
representation. Nevertheless, a good estimationof the phase
transition point can be identified by suitable scaling analysis \cite{scaling}.

Even though we can complete the finite size calculations using
finite MPS to obtain the gap for larger systems, due to the fast
convergence of this transition value for short chains this task
brings little additional information. Moreover, the precision
required to assess the $\frac{1}{N}$ dependence found above is
numerically demanding. This requires {\emph i.e.} $\chi=20$ for size
$N=12$, in order to obtain the transition for a value of $h$
compatible with the exact diagonalization. Lower values of $D$ will
place this transition at higher $h$, which exemplifies the need for
relatively high $\chi$ values even for small chains, due to the
critical character of the transition.

We can access  directly by means of the iTEBD algorithm a representation of the
infinite chain by imposing translational invariance in the tensor
description (for details see \cite{Vidal07}). The final process
however will produce energy and magnetic moment per particle, as
opposed to the total energy and total magnetic moment of the exact
diagonalization and MPS for finite systems. This means that we
cannot extrapolate the energy from the transition point back to
$h=0$ to obtain the vertical offset as the energy gap. However,
using the second viewpoint mentioned earlier, the field value at
which the energy density and the magnetization becomes negative is
the value of the energy gap, and no extrapolation is necessary.

\begin{center}
\begin{figure}[htb]
\includegraphics[width=\linewidth]{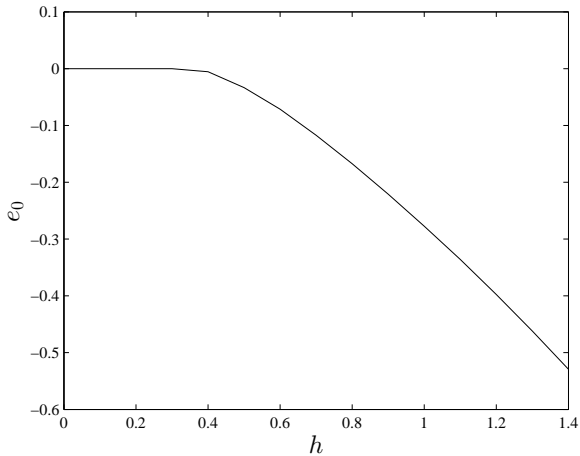}
\caption{Ground state energy per site of the AKLT Hamiltonian for an
infinite chain as computed using iTEBD with $\chi=30$.
We can identify a plateau around $h=0$, and the transition to $e_0<0$
indicates a change in the ground state. The value $h_c$ is used to obtain the
gap $\Delta$.} \label{fig:itebd_10_10_energy}
\end{figure}
\end{center}

\begin{center}
\begin{figure}[htb]
\includegraphics[width=\linewidth]{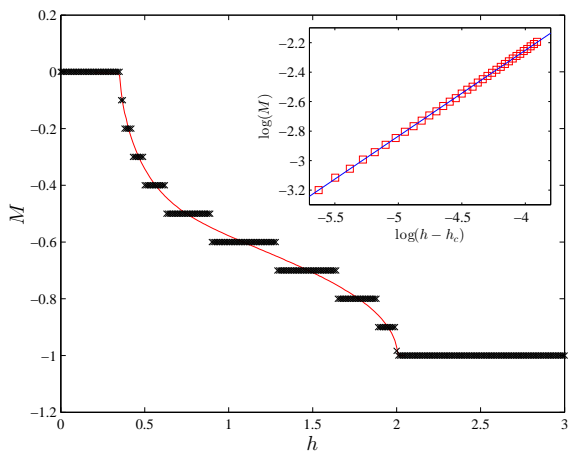}
\caption{$S_z$ per site of the AKLT Hamiltonian for an
infinite chain as computed using iTEBD with $\chi=30$ (red solid line).
The exact results for $N=10$ are shown in black crosses as a reference.
Inset: With a scaling fit of $S_z$ close to the transition, we obtain
a value of the transition $\Delta=0.350$. A bigger bond dimension
$\chi=60$ is used to perform this fit.} \label{fig:itebd_10_10}
\end{figure}
\end{center}

We show the iTEBD results for the energy per site in
Fig.\ref{fig:itebd_10_10_energy}, where we can observe clearly a
plateau up to $h\sim 0.35$, indicating the gapped phase. This result
resembles those in Fig.\ref{fig:exact_12_detail} for $N=12$, due to
the fast convergence of the spectrum. The solid line in
Fig.\ref{fig:itebd_10_10} shows $M=S_z$ per site in the
thermodynamic limit as computed using $\chi=60$ iTEBD. The energy
curve displays a plateau followed by a transition at $h=0.350$. The
points in Fig.\ref{fig:itebd_10_10} are results for $N=10$ obtained
by exact diagonalization, displayed here as a reference. For an
accurate estimation of the transition, we study the scaling $S_z\sim
(h-h_c)^\beta$ to obtain $\Delta = 0.350$. These results are
computed using $\chi=60$ and shown in the inset of
Fig.\ref{fig:itebd_10_10}.

At this point it is interesting to consider the limiting cases we
have already explored. Observing the magnetization curves, new
plateaus of the magnetization appear for increasing $N$
corresponding to the new values available to the magnetization from
the new spin combinations. However, the gap and the value of the
field for the totally polarized state are similar for different
sizes. This restricts the existing and new plateaus of the
magnetization into a the same range of field values. Progressively
with increasing $N$ the plateaus will shorten their width, and in
the limit $N\rightarrow \infty$ the plateaus will appear at any
value of $S_z$ (per particle) with an infinitesimal width. This is
clearly depicted in Fig.\ref{fig:itebd_10_10}. The transition point
--as read from the magnetization-- agrees with the value $\Delta =
0.350$ obtained from the finite size scaling.

We comment on an interesting observation. The last plateau where the
spins become completely polarized $|-1,-1,-1,...\rangle$ begins at
$h=2$. This is due to a transition from a
previous plateau characterized  by a state of the form
\begin{equation}
|\psi\rangle=\frac{1}{\sqrt{N}}\sum_{j=1}^N (-1)^{j}
|-1\rangle|-1\rangle...|-1\rangle_{j-1} |0\rangle_j
|-1\rangle_{j+1}...|-1\rangle.
\end{equation}
By direct calculation, we can show that this is an eigenstate and
has energy $(N-2)-h (N-1)$. Comparing with the completely polarized
state, which has energy $N- h N$, we find that the crossing occurs
at the $h=2$. As we shall see in the two dimensional case, the same
type of states give the last crossing at $h=3$ in the honeycomb and
$h=4$ in the square lattice case (in general $h=z$, the number of
neighbors).

The existence of a finite nonzero gap in the thermodynamic limit was
already established by AKLT~\cite{AKLT2} via lower bounds
and their technique was subsequently
generalized by Knabe~\cite{Knabe} and by Nachtergaele~\cite{Nachtergaele}.
However, the lower bounds provided by these methods are not tight. Our
method for accessing the gap via the external field $h=\Delta$ holds
even for $N\rightarrow \infty$ whenever the first excited state --
which is a triplet-- has total $S_z=-1$ (as the system approaches
infinity\cite{affleckPRB2}). In this direction is important to study the structure for
the first excited state. In \cite{Knabe} the elementary excitations
(so-called {\it crackions\/}) are expressed as a variational
perturbation over the AKLT state, by the superposition of states
formed by breaking one of the virtual singlet bonds. These
excitations in fact are not eigenstates of the Hamiltonian in
Eq.~\ref{eqn:1DAKLT} but provide an upper bound of the energy of the
first excited state. This value $\Delta=\frac{10}{27} \cong 0.37$ can
also be derived from the single-mode approximation~\cite{Auerbach}
and is close to but higher than our estimation $\Delta = 0.350$.

From the proposed excited states  we also discard the existence of
other states with $S^{tot}=0$ between the ground state and the first
excitations with $S^{tot}=1$. These excitations don't interact with the
field $h$ and thus will never cross the ground state, making them
invisible to our method. However, field-theory arguments\cite{affleckPRB}
and numerical results \cite{whiteGap,HeisGap} indicate that the first excitation is a triplet
with $S^{tot}=1$, and the first excitation with $S^{tot} = 0$ appears
with very high energy. This is consistent with our finite size
results of Table \ref{fig:table_gap} and Fig. \ref{fig:finite_gap_scal}.

\subsection{1D Bilinear biquadratic Model} \label{section:bilinear}

The AKLT model is a particular case of the family of the bilinear
biquadratic Hamiltonian written as \be \label{eq:bilinear} H_{BB} =
\sum_{\langle i,j\rangle} \left[ \vec{S}_i\cdot \vec{S}_j +
\gamma(\vec{S}_i\cdot\vec{S}_j)^2 \right]. \ee We note that the AKLT
model corresponds to $\gamma=1/3$. However, in order to write the
AKLT as a composition of projecting operators one includes an
overall factor $J=1/2$ (see Eq.~(\ref{eqn:1DAKLT}) as the
Hamiltonian used earlier).

As a direct extension of the previous analysis of the AKLT model we
use the same techniques to explore the gap of Eq.~\ref{eq:bilinear}
for the Heisenberg model, \emph{i.e.} the case with $\gamma = 0$.
Previous DMRG calculations show accurately a gapped phase with
$\Delta=0.4107$~\cite{whiteGap}, and bounds to this gap have been
calculated in \cite{HeisGap}. In Ref.\cite{affleckPRL} a bosonic
model for the excitations of the Hesenberg model provides a scaling
function around the transition to $e_0<0$ as \be M\sim
\frac{\sqrt{(h-h_c)\Delta}}{2v\pi}, \ee where $\Delta$ is the energy
gap and $v$ the magnon velocity.

\begin{center}
\begin{figure}[htb]
\includegraphics[width=\linewidth]{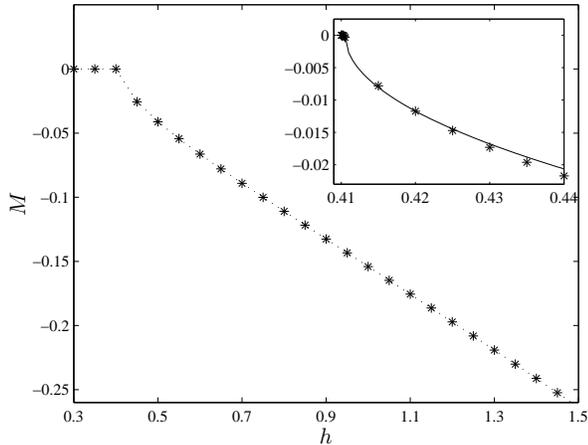}
\caption{For the Heisenberg spin-1 model, the transition in the
magnetization $M$ vs. the field $h$. Excitations are states of
$S_z^{tot} = -1$, and we read directly the gap from the transition
in the ground state magnetization $S_z^{tot}\neq 0$. The dotted line
is included as a reference, not actual data. Inset: Detailed view of
the transition region. The solid line is a fit to the function
$M\sim \frac{\sqrt{(h-h_c)\Delta}}{2v \pi}$. From the fit we obtain
$\Delta=0.4105$ and $v=2.37$.} \label{fig:heisenberg}
\end{figure}
\end{center}

Using the method presented above we compute the gap and magnon
velocity using a fit of $S_z$ around $h_c$, which also provides the
value of the gap $\Delta$. Our results are shown in
Fig.\ref{fig:heisenberg}. These results are obtained with iTEBD and
$\chi=80$ with fourth order Trotter evolution for the state
preparation. From the fit (see the inset of
Fig.\ref{fig:heisenberg}) we obtain $\Delta=0.4105$ and $v=2.37$, in
good agreement with the results in \cite{affleckPRL,WeyrauchRakov}
and the DMRG result in \cite{whiteGap}.

\section{Hexagonal lattice in 2D} \label{section:hexagonal}

Using PBC, the spin-3/2 AKLT state, e.g, on the honeycomb lattice is
the unique ground state of the following Hamiltonian with spin
rotational symmetry
\begin{eqnarray} \label{eqn:spin3/2AKLT}
 H^{S=3/2}_{\rm AKLT}
&=&\frac{27}{160}\sum_{\langle i,j\rangle}\Big[ \vec{S}_i\cdot
\vec{S}_{j}+\frac{116}{243}(\vec{S}_i\cdot \vec{S}_{j})^2
\nonumber\\
&& +\frac{16}{243}(\vec{S}_i\cdot \vec{S}_{j})^3 +
\frac{55}{108}\Big],
\end{eqnarray}
defined on trivalent lattices~\cite{AKLT2}. If open boundary
conditions are used or the boundary spin-3/2's are terminated by
spin-1/2's with Heisenberg-type interaction (between the spin-1/2
and the boundary spin-3/2)  the ground state is unique. Similarly,
AKLT states defined on any tetravalent lattice, such as the square,
Kagom\'e and the 3D diamond lattices, are the ground states of a
spin isotropic Hamiltonian with the highest order term proportional
to $(\vec{S}_i\cdot \vec{S}_{j})^4$.

In order to study this lattice in the thermodynamic limit
we use the TNRG\cite{levin,wen,xiangPRB} method to obtain expectation
values of the energy and magnetization per site. This method is governed by a parameter
$D$ for the state preparation by local update, and $D_{cut}$ for the RG calculation.
Obtaining values for
finite systems with TN requires either setting PBC or including
boundary spin-1/2 particles to obtain a unique ground state. These
two approaches are more involved than TNRG, and we are mainly
interested in proving propertied of the infinite lattice. For
exact calculations in finite systems, we impose PBC in our numerics.

The approach to 2D systems presented here is the same as for the 1D spin chain.
In the infinite limit we have only access to expectation values per
site, which is enough to identify the transition between the ground
state and excited states at some $h$, as will be shown. Scanning for different values
of $h$ and computing the energy and magnetization, we identify the
transition at which they first become negative, i.e., $e_0<0$ and
$m_z <0$.

\begin{center}
\begin{figure}[htb]
\includegraphics[width=\linewidth]{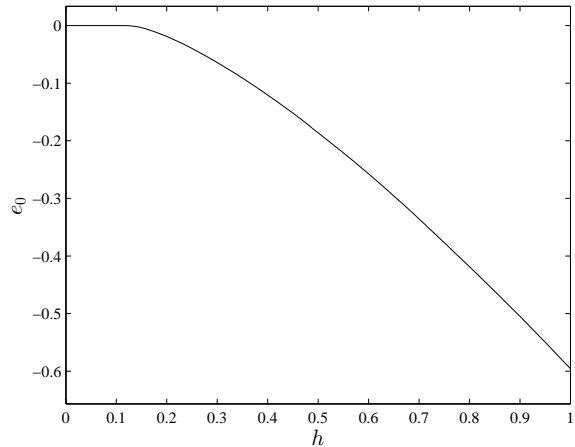}
\caption{Ground state energy per site $e_0$ as a function of the
field $h$, computed with $D=3$ and $D_{cut}=20$. We observe a flat
region around $h=0$ indicating the presence of a gap $\Delta$. The
transition to $e_0<0$ at $h_c$ provides a direct reading of a lower
bound to the gap.} \label{fig:inf_hex_energy}
\end{figure}
\end{center}

\begin{center}
\begin{figure}[htb]
\includegraphics[width=\linewidth]{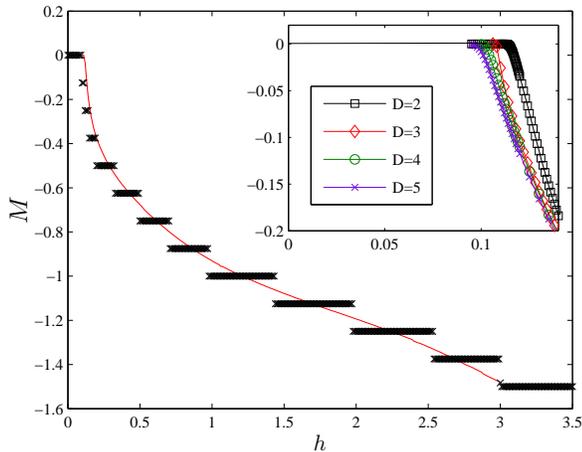}
\caption{$S_z$ per site of the AKLT Hamiltonian
for an infinite hexagonal lattice as computed using TNRG with $D=3$
and $D_{cut} = 20$ (red solid line). The cross marks are exact
results for a lattice $2\times 4$ using PBC.
Inset: Transition in the magnetization of the infinite
hexagonal lattice using $D=2,3,4,5$ using $D_{cut}=20$.}
\label{fig:inf_hex}
\end{figure}
\end{center}

We plot the results of this procedure to obtain the energy per site in Fig.\ref{fig:inf_hex_energy}
using $D=3$ for the ground state tensors, and $D_{cut}=20$ for the RG method.
We clearly observe a plateau of $e_0=0$ up to the value $h_c\sim0.1$. A more accurate analysis
is presented for the magnetization $S_z$ per site in Fig.\ref{fig:inf_hex}.
We again observe a flat region before
the transition at $h_c\sim0.1$, indicating the presence of an energy gap.
These results are in good agreement with finite size calculations \cite{sylvain,ganesh} over
small clusters of spins.
The inset shows results around the transition point using $D=2,3,4,5$ for the ground state tensors,
and $D_{cut}=20$ for the RG method. As a remark, we observe the transition to the fully polarized state
of the hexagonal lattice at a value of the field $h_f=3$. As explained above,
this transition happens at $h_f = 2s$ (with $s$ the local spin).

\begin{center}
\begin{figure}[htb]
\includegraphics[width=0.7\linewidth]{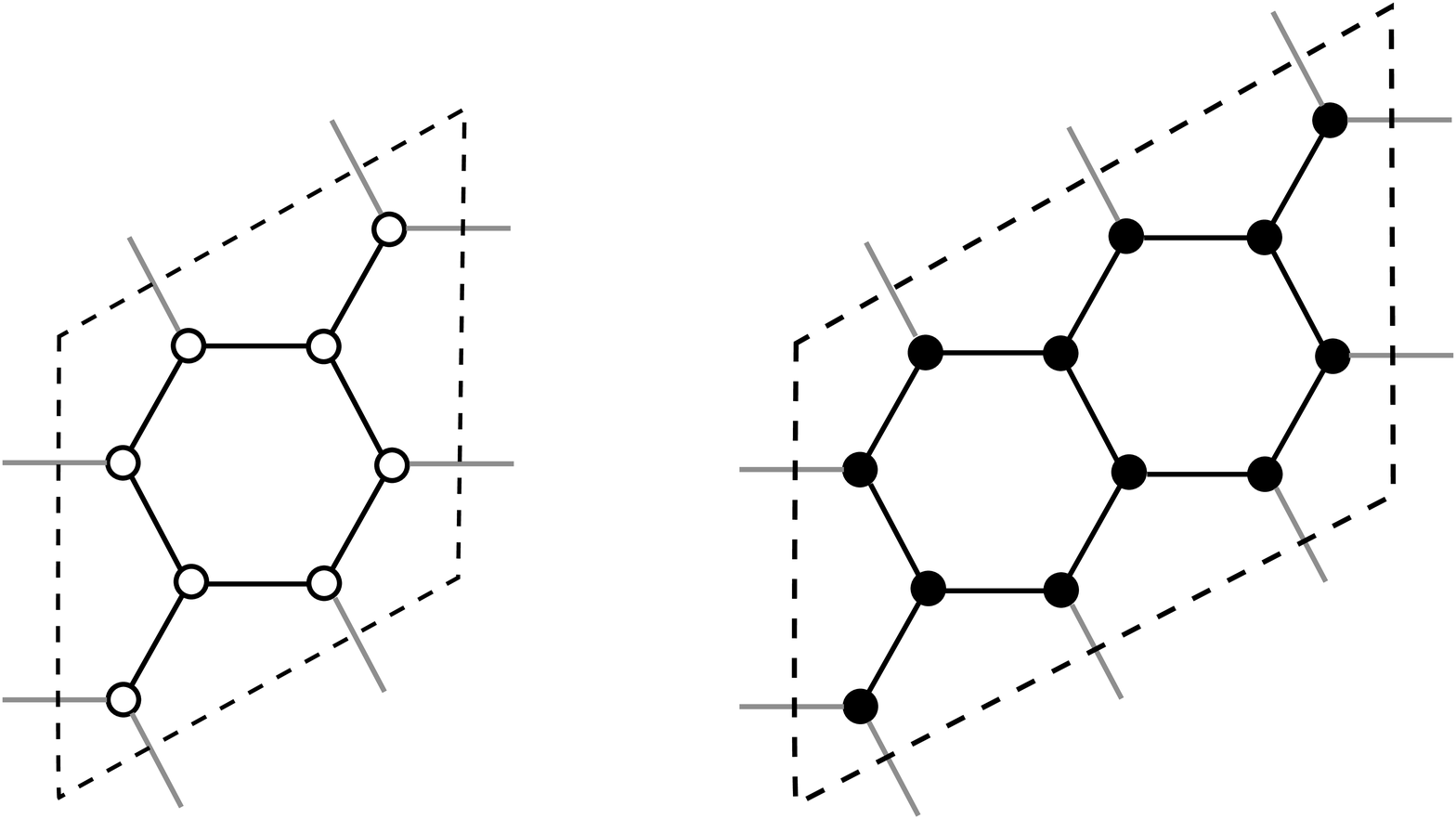}
\caption{Disposition of the lattices used for finite size calculations
of the finite hexagonal lattice with PBC to compute the gap
by exact diagonalization. For a lattice (white circles, left) we find
$\Delta_{2\times4} = 0.092029$. Introducting $4$ additional sites (to the right in black circles,
with a total size $3\times 4$) we obtain $\Delta_{3\times4} = 0.095345$.}\label{fig:table_hex}
\end{figure}
\end{center}

We compare the TNRG results with exact diagonalization for small
lattices as a reference. We construct a small lattice with PBC of size $2\times 4$
and $3\times 4$ to obtain the gap (see Fig.~\ref{fig:table_hex}).
For these settings we obtain the gap values
$\Delta_{2\times 4} = 0.092029$ and $\Delta_{3\times 4} = 0.095345$.
Notice that again the first transition provides an
exact reading of the gap: this suggests that for small systems the single spin
excitations are the lowest excited states, as in the 1D chain. However,
without bounds for the excitation energy in the infinite hexagonal lattice,
we can only conjecture that the transition values coincide with the gap in the
thermodynamic limit, \emph{i.e.} the elementary excitations have
$S^{tot}=1$.

As shown in the inset of Fig.~\ref{fig:inf_hex} the transition point
shifts to lower values of $h_c$ as the bond dimension $D$ increases.
Exact diagonalization results suggest --as in 1D chains-- higher
values of the gap for larger systems. Identifying a critical
transition using Tensor Network methods requires large bond
dimensions and a careful choice of the interval for the scaling
fit\cite{sandvik}.

\begin{center}
\begin{figure}[htb]
\includegraphics[width=\linewidth]{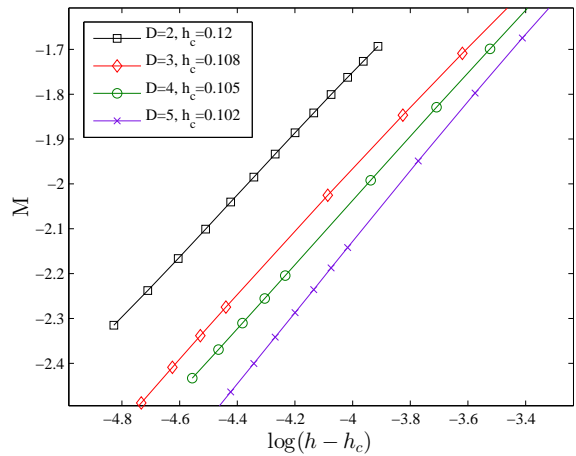}
\caption{Using the scaling relation $M \sim (h-h_c)^\beta$
we obtain the value of the critical transition
for $D=2,3,4,5$ at $h_c=0.12, 0.108, 0.105, 0.101$ respectively.
Using the energy instead of the magnetization (not shown here),
we obtain similar transition values.}\label{fig:scaling_hex}
\end{figure}
\end{center}

To identify the infinite-system
transition point from our results using TNRG (see the inset in Fig.\ref{fig:inf_hex}),
we perform a scaling analysis for different values of $D$.
We use the scaling relation for the magnetization
close to the transition point $M \sim(h-h_c)^\beta$.
Fitting the energy, we obtain $h_c = {0.12,0.108,0.104,0.100}$
for $D=2,3,4,5$ respectively. Using the magnetization in a similar way
we obtain $h_c={0.12,0.108,0.105,0.101}$. The fitting results
for the magnetization $S_z$ are shown in Fig.\ref{fig:scaling_hex}.

\section{Square lattice in 2D: infinite study with TNRG} \label{section:square}

We finally study the 2D square lattice, where the AKLT state can be
viewed as 4 spin-1/2 particles at each site, for a
total local spin $s=2$. The AKLT hamiltonian projects two neighboring $s=2$ particles
into the spin $s=4$ subspace, resulting in the Hamiltonian
\begin{eqnarray} \label{eqn:spin2AKLT}
 H^{S=2}_{\rm AKLT}
&=&\frac{1}{14}\sum_{\langle i,j\rangle}\Big[ \vec{S}_i\cdot
\vec{S}_{j}+\frac{7}{10}(\vec{S}_i\cdot \vec{S}_{j})^2
\nonumber\\
&& +\frac{7}{45}(\vec{S}_i\cdot \vec{S}_{j})^3 +
\frac{1}{90}(\vec{S}_i\cdot \vec{S}_{j})^4\Big].
\end{eqnarray}

\begin{center}
\begin{figure}[htb]
\includegraphics[width=\linewidth]{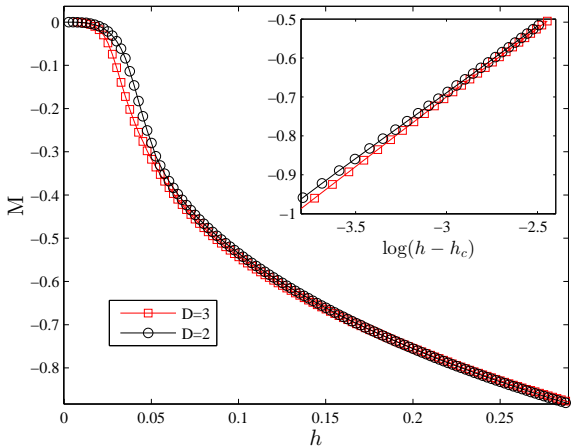}
\caption{Magnetization curve for the spin-2 AKLT Hamiltonian in
presence on an external field $h$ on the square lattice.
We show here results for $D=2,3$
simulations using TNRG where we set $D_{cut}=20$.
Inset: Using the scaling reation $M \sim(h-h_c)^\beta$ we
plot $\log(h-h_c)$ vs. $\log(M)$, from where we obtain a transition
at $h_c=0.03$.}
\label{fig:square_2}
\end{figure}
\end{center}

Following the same procedure as for the spin-1 and spin-3/2 AKLT
systems, we compute the transition $h_c$ from the magnetization
$S_z$.  Our results are obtained solely using TNRG in the infinite
limit, and are shown in Fig.~\ref{fig:square_2} for $D=2,3$. Even
though the plateau cannot be clearly observed as in the other
models, a scaling fit of the magnetization (see the inset in
Fig.~\ref{fig:square_2}) is compatible with a gap value
$\Delta=0.03$. Such a small value has to be considered together with
the prefactor $J=\frac{1}{14}$ so the Hamiltonian is formed as a
combination of projecting operators.

\section{Conclusions} \label{section:conclusions}

The study of the gap of AKLT Hamiltonians conducted here starts with
exact diagonalization for finite systems that suggest the existence
of a gapped phase and provides evidence supporting that the first
excited states form a triplet. For 1D spin-1 chains we have both
lower and upper bounds for this gap, which agree with our finite
size scaling results. Using MPS we can access longer systems sizes,
but the gap is clearly observed to converge quickly even at small
chain lengths. We complete the picture in 1D with iTEBD results that
also agree with the finite size results, resulting in a gap for the
spin-1 AKLT chain of $\Delta\cong0.350$.

In 2D, numerical exact results reach only hexagonal lattices of
small size. Finite size PEPS simulations may complement these
results for a proper finite size scaling; however, we did do this.
Instead, using TNRG techniques we directly access the thermodynamic
limit and obtain a value of the gap $\Delta\cong0.10$ in agreement
with the finite results. This agreement arises from two
observations: the gap increases with the system size, and the fast
convergence of the gap with $N$ can be observed even in very small
systems. For the square lattice we cannot obtain exact results for
appropriate sizes and we rely solely on TNRG results. These results
suggest also a gapped phase with $\Delta\cong0.03$.

Our method to compute the gap in the limit of infinite system size
introduces an additional field term in the Hamiltonian commuting
with it and this is equivalent to probing the first quantum phase
transition as $h$ increases from $0$. We have shown results for the
AKLT Hamiltonian, and in general for the bilinear biquadratic model
in the gapped phase. Our results have good agreement with previous
bounds and DMRG result for 1D spin chains. Nevertheless, the general
idea of this method can be applied by identifying symmetries in
ground states of Hamiltonians commuting with the additional external
field. It would be desirable to have analytic proof for the
existence of the gap.

Note added: Similar results on the honeycomb lattice were recently obtained in \cite{psc}.

\acknowledgments The authors acknowledge fruitful discussions with
Robert Raussendorf, Oliver Buerschaper, Andreas L\"auchli,  Frank
Verstraete, and especially Sylvain Capponi.

\end{document}